\documentclass[twoside,11pt]{amsart} 

\usepackage{amsfonts}
\usepackage{amssymb}
\usepackage{amsmath}
\usepackage{graphicx}

\usepackage[all]{xy}
\usepackage{mathrsfs}

\def\noi{\noindent}

\newtheorem{thm}{Theorem}[section]

\newtheorem{conj}[thm]{Conjecture}

\newcommand{\mathset}[1]{{\left\{#1\right\}}} 
\newcommand{\absolute}[1]{\lvert#1\rvert}

\DeclareMathOperator{\Aut}{Aut}

\DeclareMathOperator{\Val}{Val}
\DeclareMathOperator{\trop}{trop}

\DeclareMathOperator{\Mumf}{Mumf}
\DeclareMathOperator{\PGL}{PGL}

\begin{document}
 \baselineskip=11pt

\title{Counting plane Mumford curves}
\author{{Patrick Erik Bradley}}


\date{}

\maketitle

\begin{abstract}
\noi
A $p$-adic version of Gromov-Witten invariants for counting
plane curves  of genus $g$ and degree $d$ through a
given number of points is discussed. 
The multiloop version of $p$-adic string theory considered by 
Chekhov and others
motivates us to ask how many of these curves are Mumford curves, i.e.\
uniformisable by a domain at the boundary of the Bruhat-Tits tree
for $\PGL_2(\mathbb{Q}_p)$.
Generally,  the number of Mumford curves
depends on the position of the given points in $\mathbb{P}^2$.
With the help of tropical geometry we find configurations 
of points through which all curves 
of given degree and genus are Mumford curves.
The article is preceded by an introduction to some concepts of
 $p$-adic geometry and their relation to string theory.
\end{abstract}




\section{Introduction}

Since the work of Volovich \cite{Volovich} string theory has profited from $p$-adic methods.
However, each $p$-adic field $K$ has its own string theory.
The consideration of classical string theory as a limit
of $p$-adic string theories for ``$p\to 1$'' requires
a unified approach for all $p$-adic number fields for fixed prime number $p$.

We propose $p$-adic geometry as a framework for realising this task.
In this article, we introduce methods from this framework with 
string theoretic relevance. Some of these have been applied
to the analysis of hierarchical data \cite{BradJoC}. More
methods are developped in $p$-adic enumerative geometry \cite{Brad}.
Of particular interest are the Mumford curves which play a role in
the $p$-adic multiloop calculations in \cite{CMZ1989}.
Conjecturally, these special curves are the only ones
contributing to the string amplitude \cite[Conj.\ 4.3.3]{CMZ1989}.
From the point of view of so-called {\em  tropical geometry}, 
this CMZ-conjecture, as we call it,
 should come
natural. The reason is that tropical curves
are generically obtained from transforming
 Mumford curves into combinatorial objects.
In any case, our work is motivated by the conjecture.

The aim of our methodological overview is twofold.
Primarily, we want to show how they can be used to count plane Mumford curves.
Secondly, we indicate how the methods could give a
positive answer to a more precise formulation of the CMZ-conjecture. 
Our long-term goal is to be able to ``predict'' enumerative results
for Mumford curves with $p$-adic string theory, similarly as in the classical case---the only
difference being that the mathematical answers might be known before their
physical derivations.

We refer to the article by Dragovich
\cite{Dragovich} for an introduction to $p$-adic numbers
and their relation to string theory.

\section{Pr\'elude: An introduction to $p$-adic geometry}

Let $\mathbb{Q}_p$ be the field of $p$-adic numbers.
In the following, we will often use the notation $\absolute{x}$
for $\absolute{x}_K$, where $K$ is any finite extension field of $\mathbb{Q}_p$
containing $x$. This notation is well defined. In fact, we could as well
consider $x$ as an element of $\mathbb{C}_p$,
 the completion of the algebraic closure of $\mathbb{Q}_p$,
and $\absolute{\;}=\absolute{\;}_{\mathbb{C}_p}$,
 the unique extension of $\absolute{\;}_p$
to $\mathbb{C}_p$. By $O_K$, we denote
the ring of integers
$$
O_K=\mathset{x\in K\mid \absolute{x}_K\le 1},
$$
and $\kappa=O_K/\pi O_K$ is the residue field. It
is finite and does not depend
on the choice of the {\em uniformiser} $\pi$ which generates
the maximal ideal of $O_K$. 

The field $K$ has an affine geometry. Hence, we can write
$K=\mathbb{A}^1(K)$. However, this space is only the set of $K$-rational
points of the geometric object $\mathbb{A}^1$ which we call {\em affine line}.
We will often make the distinction between a space $X$ and its
$K$-rational points $X(K)$.

The topology of $p$-adic spaces such as $\mathbb{A}^1$ is
totally disconnected. This uncomfortable fact can be remedied e.g.\
by introducing extra points. Here, we do this with the method from
\cite{Berkovich1990} 
and call the extra points {\em Berkovich points}.
In the example of the  affine line $\mathbb{A}^1$,
the important Berkovich points correspond to
 the discs $B_a=\mathset{\absolute{x-a}\le r}$
with $r>0$.

\subsection{Projective spaces}

The idea of projective space is to have a 
good compactification of affine space
which is locally affine. {\em Projective $n$-space} over $K$ is
$$
\mathbb{P}^n(K):=\mathset{\text{lines through $0\in K^{n+1}$}}
$$
One has a decomposition 
$\mathbb{P}^n(K)=\mathbb{A}^n\cup\mathbb{P}^{n-1}(K)$,
i.e.\ another projective space ``at infinity''.
Projective coordinates are often written as 
$$
(x_0:\dots:x_n)
$$
with $(x_0:\dots:x_n)=(y_0:\dots:y_n)$ if and only if
there is some $\lambda\neq 0$ such that
$x_i=\lambda y_i$ for all $i$.
The local structure is given by
$$
\mathbb{P}^n=U_0\cup\dots\cup U_n
$$
with affine pieces 
$$
U_i=\mathset{\left(\frac{x_0}{x_i},\dots,\frac{x_n}{x_i}\right)\mid x_i\neq 0}
\cong\mathbb{A}^n.
$$
For example, $\mathbb{P}^1=\mathbb{A}^1\cup\mathset{\infty}$
is the projective line.
The projective plane is $\mathbb{P}^2=\mathbb{A}^2\cup\mathbb{P}^1$. It
has the property that any two lines in $\mathbb{P}^2$ intersect.

The space $\mathbb{P}^n$ is endowed in a natural way with a line bundle.
Namely, for $x\in \mathbb{P}^n(K)$ let $\ell_x$ be the line in $K^{n+1}$
represented by the point $x$. This line bundle is the {\em tautological line
bundle} $O(1)$ encountered later on.

\subsection{Bruhat-Tits tree}

\begin{figure}[h]
$$
\underbrace{
\hspace*{-10mm}
\xymatrix@=20pt{
&\ar@{-}[dd]&\ar@{-}[dd]&&\ar@{-}[dd]&\\
\ar@{-}[rrrrr]&&&&&\\
&&&\dots&
}
\hspace*{-10mm}
}_{q+1}
$$
\caption{Tree of projective lines.} \label{infproj}
\end{figure}
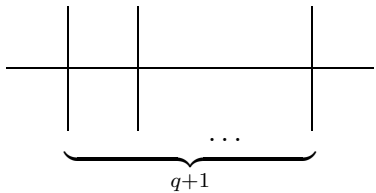

The symmetry group of the projective line $\mathbb{P}^1$ 
over $K$ is $\PGL_2(K)$, the group of fractional  transformations
\begin{align}
z\mapsto\frac{az+b}{cz+d} \label{Moebiustrafo}
\end{align}
with $ad-bc\neq 0$. The map (\ref{Moebiustrafo}) is also called {\em M\"obius 
transformation}.
The fact that M\"obius transformations take discs to discs
allows to construct an infinite tree
$\mathscr{T}_K$ on which $\PGL_2(K)$
acts as group of symmetries. This tree is the {\em Bruhat-Tits tree}
for $\PGL_2(K)$ and can be visualised as the hierarchical tree
of discs
$$
B_a=\mathset{x\mid\absolute{x-a}_K\le \absolute{r}_K},
$$
the vertices being given by $B_a$ and an edge  is given by
maximal strict inclusion $B_b\subset B_a$ of 
discs\footnote{However, there is some subtlety
concerning the invariance under $\PGL_2(K)$, 
wherefore the vertices are given by equivalence classes of discs
cf.\ \cite[\S 3]{BradJoC}.},
i.e.\ any $B_c$ such that $B_b\subseteq B_c\subseteq B_a$
satisfies either $B_c=B_b$ or $B_c=B_a$.
The tree $\mathscr{T}_K$
 is a $q+1$-regular tree, meaning that from each vertex
there are precisely $q+1$ edges going out, where $q$ is the
cardinality of the residue field $\kappa$.
The geometric reason behind this fact is that every vertex $v$
of $\mathscr{T}_K$
corresponds to a 
projective line $\mathbb{P}^1_v$,
and  its attached
edges correspond to the $\kappa$-rational points $\mathbb{P}^1_v(\kappa)$.
Hence, $\mathscr{T}_K$ can be seen as representing
the combinatorics of infinitely many projective lines
glued together as (locally) depicted in Figure \ref{infproj}.

An important fact is that the boundary of the tree
$\mathscr{T}_K$ is given by $\mathbb{P}^1(K)$.
Namely, an infinite path in $\mathscr{T}_K$ can be understood
as a strictly
descending sequence of discs whose limit is their intersection:
a $K$-rational point in $\mathbb{P}^1$.

\medskip
Let now $C=\mathbb{P}^1\setminus\mathset{p_1,\dots,p_n}$
be the projective line over $K$ with $n$ points
(also called {\em punctures})
$p_1,\dots,p_n$ removed. These points define a subtree 
$T=\mathscr{T}(p_1,\dots,p_n)$ of $\mathscr{T}_K$
by connecting all geodesic paths inside the Bruhat-Tits tree
between the punctures. In \cite{BradJoC}, 
this tree was interpreted as dendrogram for the ``data''
$p_1,\dots,p_n$.

The tree $T$ corresponds to the glueing of projective lines
$\mathbb{P}^1_v$ over $\kappa$ for each vertex $v$ and then
removing $n$ punctures. This geometric object is a singular
curve $C_s$, the singularities being ordinary double points,
and the lines $X_i$ constituing the curve $C_s$ are the {\em irreducible
components}. They   are represented in
$\mathbb{P}^1$ by discs corresponding to 
Berkovich points $\xi_i$.

\subsection{Mumford curves}

The $p$-adic analogon of Riemann surface in the physics literature
is the {\em Mumford curve}. It allows a Schottky uniformisation: if
$F_g$ is a discrete subgroup of $\PGL_2(K)$ which is generated by $g$ 
hyperbolic transformations, then $X=\Omega/F_g$ is a complete algebraic
curve. Here, $\Omega\subseteq\mathbb{P}^1$ is the domain of regularity
of the action of $F_g$. 

Not every $p$-adic algebraic curve allows a Schottky uniformisation.
However, there are some characterisations of Mumford curves.
Namely, every $p$-adic curve $X$ has a so-called {\em $O_K$-model}.
It is a curve $\mathcal{X}$ defined over the ring $O_K$ of integers of $K$:
consider some (local) set of equations for $X$, and clearing all
denominators yields equations with coefficients in $O_K$.
Then reducing all equations modulo $\pi$ yields a curve $X_s$ defined over 
$\kappa$, called the {\em special fibre} of $\mathcal{X}$.
In general, $X_s$ is singular, even if $X$ is not.
By a theorem of Deligne and Mumford \cite[Cor.\ 2.7]{DM1969},
it is possible for $K$ sufficiently large to find an $O_K$-model $\mathcal{X}$
such that $X_s$ is a so-called {\em stable} curve, meaning:
\begin{itemize}
\item All singularities of $X_s$ are ordinary double points.
\item $\absolute{\Aut X_s}<\infty$.
\end{itemize}
There is a reduction map 
\begin{align}
\rho\colon X\to X_s \label{redmap}
\end{align}
which is locally 
``reduction modulo $\pi$''. The upper curve $X$ is called the {\em generic
fibre} of $\mathcal{X}$.

\medskip
We now assume that $K$ is sufficiently large.
The characterising
 criterion for $X$ being a Mumford curve is then that the special
fibre $X_s$ is a union of genus zero curves \cite[Thm.\ 5.4.1, 5.5.5]{FP2004}.

The special fibre $X_s$ of a stable curve allows a combinatorial
description by taking as vertices the irreducible components
of $X_s$ and as edges the double points. The resulting graph 
$\Gamma$ is the {\em dual graph} of $X_s$.
This yields the next characterisation: $X$ is a Mumford curve, if and
only if the first Betti number of the dual
graph $\Gamma$ of $X_s$ equals the genus of $X$.

Let now $X$ be a curve with $n$ punctures. Then the dual graph of $X$
can be obtained from the dual graph $\Gamma'$ of the completion of $X$ by
adhering some infinitely long spines to $\Gamma'$.
The result is a so-called {\em $n$-pointed tropical curve}
 $\Gamma=\trop(X)$. We call the combinatorial object underlying
$\Gamma$ a {\em semigraph}, and the spines are the {\em punctures}.
There is a metric on $\Gamma$ coming from the reduction map
(\ref{redmap}). Namely, the fibre $\rho^{-1}(x)$
of a point $x\in X_s$ is either an open disc or an open annulus $A$.
The latter holds true, if and only if $x$ is a double point.
The length of an edge of $\Gamma$ is
defined as the thickness of $A$. 

\medskip
Our central mathematical object will be the {\em moduli space of $n$-pointed
genus $g$ curves} $\mathcal{M}_{g,n}$ whose points are equivalence
classes $[C,p_1,\dots,p_n]$, where $C$ is a complete curve
of genus $g$ minus $n$ punctures $p_1,\dots,p_n$. 
The moduli space is defined over the integers $\mathbb{Z}$.
Therefore, we advocate the use of $\mathcal{M}_{g,n}$ in adelic physics,
although our focus will be on $M_{g,n}=\mathcal{M}_{g,n}\otimes K$
for $K$ a sufficiently large extension of $\mathbb{Q}_p$.

\smallskip
The methods here yield a {\em tropicalisation map}
$$
\trop\colon M_{g,n}\to M_{g,n}^{\trop}, 
[C,p_1,\dots,p_n]\to [\trop(C),p_1,\dots,p_n],
$$
where $M_{g,n}^{\trop}$ is the moduli space of $n$-pointed
tropical curves of genus $\le g$. The punctures of $\trop(C)$ are labelled
in the same way as the punctures of $C$. 
The moduli spaces are not compact. The {\em Deligne-Mumford compactification}
$\bar{\mathcal{M}}_{g,n}$ is defined by including the stable curves.
This allows to define $\bar{M}_{g,n}^{\trop}$
as the space parametrising tropical curves whose
edge lengths can take any value between $0$ and $\infty$.
The latter comes from a singularity in the generic fibre $C$.
In the former case, it can happen that loops get contracted to a vertex.
Then $\trop(C)$ is not the tropicalisation of a Mumford curve,
as the Betti number is lower than the genus of $C$ (cf.\ also \cite{Brad}).

If $g=0$, then $\trop(C)$ coincides with $\mathscr{T}(p_1,\dots,p_n)$
from the previous subsection. Also the singular curve $C_s$
considered there is the special fibre of an $O_K$-model $\mathcal{C}$
of $C$.

\subsection{Tropical geometry of the $p$-adic projective plane}

We give here a very brief introduction into the aspects
of tropical geometry which we later use.
A more general introduction
to tropical geometry can be found e.g.\ in \cite{Mikhalkin2007,Mikhalkin}.

\medskip
The valuation map
$$
\Val\colon (K\setminus\mathset{0})^2\to\mathbb{R}^2,
\;(x,y)\mapsto (v_K(x),v_K(y)),
$$
with $v_K(z)=-\log\absolute{z}_K$,
has as its image the lattice $\frac{1}{e}\mathbb{Z}^2$ in the Euclidean
plane, where $e$ is the ramification index of $K$ over $\mathbb{Q}_p$.
Making the $p$-adic field $K$ larger results in a refinement of the
lattice. In the limit, or if $K=\mathbb{C}_p$, we
obtain the rational points of the Euclidean plane.

The valuation map extends to the projective plane:
\begin{align}
\Val\colon \mathbb{P}^2\to\mathbb{TP}^2 \label{val}
\end{align}
by defining $v_K(0)=\infty$
 on each affine patch $U$, i.e.\
we get the extra points $(\infty,y),(x,\infty),(\infty,\infty)$
on the closure of $\Val(U)$. The tropical projective
plane $\mathbb{TP}^2$ is by definition the glueing of these
closed sets. The result is  homeomorphic
 to the $2$-simplex  whose  interior corresponds to 
$\mathbb{R}^2$, and whose boundary segments correspond to
the parts with a coordinate $\infty$. 
The simplex structure reflects the fact that the complement of
$(K\setminus\mathset{0})^2$ in $\mathbb{P}^2(K)$
is the union $\mathcal{H}$
of three lines not intersecting in a common point.

\begin{figure}[h]
$$
\xymatrix@=10pt{
\scriptstyle v(y)\!\!\!\!\!\!\!\!&&&\ar@{-}[dd]&\\
&&&&\\
&&&*\txt{}\ar@{-}[dddlll]\ar@{-}[rr]&&\\
\ar[rrrrr]_{\qquad\qquad\qquad\qquad v(x)}&&&&&\\
&&&&\\
&\ar[uuuuu]&&&
}
$$
\caption{A tropical line in the plane.} \label{tropline}
\end{figure}
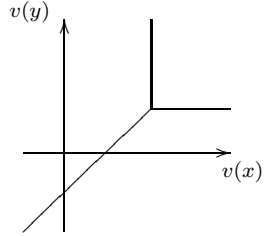

\smallskip
However, the valuation map brings more changes. It transforms $p$-adic 
geometry
to so-called {\em tropical geometry}, in which the objects
are piecewise affine-linear spaces. For example, curves
in $(K\setminus\mathset{0})^2$ transform to sets whose closures are
tropical curves embedded in the plane \cite[Thm.\ 2.1.1]{EKL2006}.
A tropical line in the plane is depicted in Figure \ref{tropline}. 
The three unbounded edges are explained by the fact that 
any line in $\mathbb{P}^2(K)$ intersects $\mathcal{H}$ in
three points. More generally, any plane curve $C$ of degree
$d$ intersects $\mathcal{H}$ in $3d$ points. This means
that the closure of $\Val(C)$ in $\mathbb{R}^2$ is a tropical
curve with
$3d$ ends (counted with multiplicity).

\medskip
One successful application of tropical geometry was
in providing elementary proofs to classical enumeration problems
of algebraic geometry. E.g.\ 
the  Kontsevich formula \cite[Claim 5.2.1]{KM1994}
 for counting  rational curves
of degree $d$ through $3d-1$ points in the plane
was obtained by counting plane tropical curves of genus zero
\cite{GathmannMarkwigWDVV}. 

\subsection{$p$-adic vs.\ tropical integration over $p$-adic spaces}\label{smoothy}

Here, we want to relate two ways of integrating over $p$-adic spaces.
The first one using the Haar measure on locally compact fields
will be called {\em $p$-adic integration}. The other method,
which we call {\em tropical integration}, is by
taking  a limit of measures coming from $p$-adic line bundles.
This allows to compute integrals via tropicalisation.

\paragraph{$p$-adic integration.}
On the locally compact additive group $\mathbb{Q}_p$, 
there is a translation invariant  measure $dx$
called {\em Haar measure}. It is usually normalised 
such that
$$
\int_{\mathbb{Z}_p}dx=1.
$$
This measure can be extended to a measure on $\mathbb{P}^1(\mathbb{Q}_p)$
by using the substitution 
\begin{align}
\phi\colon x\mapsto\frac{1}{px}, \label{substitute}
\end{align}
which changes $dx$ to $p\frac{dx}{\absolute{x}_p^2}$.
We also denote it as $dx$ and obtain
$$
\int\limits_{\mathbb{P}^1(\mathbb{Q}_p)}dx
=\int\limits_{\mathbb{Z}_p}dx
+\int\limits_{\mathset{x\in\mathbb{Q}_p\mid\absolute{x}>1}}\!\!\!dx
=1+p\int\limits_{\mathbb{Z}_p}\frac{dx}{\absolute{x}_p^2}
=1+p\cdot\frac{1}{p}=2.
$$
The same holds true with any finite extension field $K$ of $\mathbb{Q}_p$,
as long as the Haar measure $dx$ is normalised to 
$$
\int_{O_K}dx=1,
$$
where $O_K=\mathset{x\in K\mid \absolute{x}_K\le 1}$
is the ring of integers of $K$.

However, if we want to allow $K$ to vary arbitrarily among
the finite extension fields of $\mathbb{Q}_p$, then
it is often convenient to
consider $K=\mathbb{C}_p$, the completion of
the algebraic closure of $\mathbb{Q}_p$.
This approach gives some meaning to the 
limiting process ``$p\to 1$'' as explained in
\cite{Ghoshal2006},
where it is viewed as
taking a sequence of uniformisers $\pi_K$ for each $K$.
These have the property
$$
\lim\limits_K\absolute{\pi_K}_K
=\lim\limits_{e\to\infty}p^{-\frac{1}{e}}= 1, 
$$
where $e$ is the ramification index of $K$ over $\mathbb{Q}_p$.
In any case, one arrives at trying to integrate over a field which
is no longer locally compact.

\paragraph{Tropical integration.}

In order to be able to integrate over  a $p$-adic space $X$,
as opposed to its set of  $K$-rational points $X(K)$,
we use the method of Chambert-Loir \cite{C-L2006}
for $p$-adic line bundles.

\begin{figure}[h]
$$
\xymatrix@=50pt{
\ar@{-}[dr]&\ar@{-}[dl]\\
&
}
$$
\caption{A possible reduction of $\mathbb{P}^1$.}
\label{cross}
\end{figure}
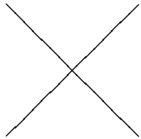

To the tautological line bundle $O(1)$ on 
$p$-adic $\mathbb{P}^1$
 can be associated a curvature form $c_1(\bar{O}(1))$
 as follows\footnote{The notation $\bar{O}(1)$ or $\bar{L}$
stands for metrised line bundle. But we suppress the definition of the
metric on $L$ for the curvature form here.}:
Let $\mathcal{X}$ be the union of two copies of projective lines
over $\kappa$ intersecting in one point as in Figure \ref{cross}.
It can be realised as a reduction modulo $\pi$ of $\mathbb{P}^1$,
and its dual graph has this shape:
$\xymatrix{*\txt{$\bullet$}\ar@{-}[r]&*\txt{$\bullet$}
}$. This corresponds to the substitution
(\ref{substitute}), and the pre-image of the double point
(represented by the open line segment) under the reduction map
$\rho\colon\mathbb{P}^1\to \mathcal{X}$ is the 
interior of the overlap $\mathbb{D}\cap\phi(\mathbb{D})$,
where $\mathbb{D}$ is the $p$-adic unit disc.
Let now $\mathcal{L}$ be the line bundle over $\mathcal{X}$ 
which is the tautological
line bundle over each component. It can be seen as a reduction of
$O(1)\otimes O(1)=O(1)^2$ on $\mathbb{P}^1$.
Then, using the algebraic version of curvature form on $X$, let
$$
c_1(\bar{O}(1)):=\frac{1}{2}\left(c_1(\mathcal{L}|_{X_1})\delta_{\xi_1}
+c_1(\mathcal{L}|_{X_2})\delta_{\xi_2}\right),
$$
where $\delta_{\xi_i}$ is the Dirac measure supported on
the Berkovich point $\xi_i$ corresponding to the 
component $X_i$. 
This
defines a Borel measure on $\mathbb{P}^1$ which induces via
 $\rho$
the measure which distributes the weight $\frac{1}{2}$ onto each
endpoint of the 
unit interval.
A careful application of
a smoothing process developped by Gubler \cite{Gubler2007}
yields a measure $\mu$ on $\mathbb{P}^1$ 
which via 
the tropicalisation map
$$
\mathbb{P}^1\to\mathbb{TP}^1,\quad x\mapsto -\log\absolute{x}_K,
$$
induces
the Lebesgue measure $d\lambda$ restricted to $\mathbb{TP}^1$ satisfying
$$
\int\limits_{\mathbb{P}^1}\mu=\int\limits_{\mathbb{TP}^1}d\lambda=1
$$
\cite{Brad}.
This
 measure
$\mu$ 
differs on $K$ from $p$-adic $dx$ only by a factor $2$. 
However, $\mu$ has the advantage that it is well-defined
over $\mathbb{C}_p$. Hence, we arrive at a tropical
interpretation of the limit ``$p\to 1$''.
We will also call $\mu$ the {\em tropical limit} of $dx$.

\medskip
Let now $X$ be a $p$-adic manifold of dimension $d$.
The generalisation of the method above needs $d$ line bundles
$L_1,\dots, L_d$ on $X$, and one obtains a regular Borel
measure by the formula
\begin{align}
\mu=c_1(\bar{L}_1)\wedge\dots\wedge c_1(\bar{L}_d)
=\sum\limits_Yc_1(\mathcal{L}_1|_Y)\wedge\dots\wedge c_1(\mathcal{L}_d|_Y)
\delta_{\xi_Y}, \label{CL-measure}
\end{align}
where $Y$ runs through the irreducible components of the special fibre of a
given $O_K$-model of $X$,
and $\mathcal{L}_i$ are specialisations of $O_K$-models of $L_i$.



\section{The $p$-adic tree-level amplitudes} \label{ptreeamp}
\begin{figure}[h]
$$
\xymatrix@=5pt{
&&&&&&\\
&\,\!+&&&&\!+&\\
&&\,\!+&&\!\!\!+&&\\
&&&&&&\\
&&\ar@{-}[uuuurrrr]&&\ar@{-}[uuuullll]&\\
}
$$
\caption{A stable $4$-pointed genus $0$ curve.}
\label{stable4dendro}
\end{figure}
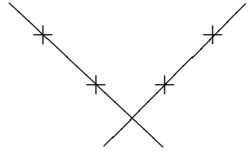

This section serves as a physical motivation for counting plane Mumford curves.
The methods from the previous section are applied to the string amplitude
at the tree level,
and will lead in the following section to an interpretation of a conjecture by Chekhov et al.\
\cite[Conj.\ 4.3.3]{CMZ1989} using more precise terms.
The named authors admittedly remained vague in formulating their conjecture.
Let us recall from  \cite{Dragovich,Volovich}
 the $p$-adic $4$-point Veneziano string amplitude 
\begin{align}
A_p^0(k_1,k_2,k_3,k_4)
=\int_{\mathbb{Q}_p}\absolute{x}_p^{k_1\cdot k_2}
\absolute{1-x}_p^{k_1\cdot k_3}\;dx, \label{4ptstringamp}
\end{align}
where $k_i\in \mathbb{C}^d$, $\sum_{i=1}^4k_i=0$
and $k_i^2=2$. Adding the point $\infty$ does not change the value of the integral, but yields a compact domain of integration $\mathbb{P}^1(\mathbb{Q}_p)$.
As in the previous section, we integrate over the space $\mathbb{P}^1$,
but now view it as a  moduli space:
\begin{align}
\mathbb{P}^1=\bar{M}_{0,4}, \label{P1=M04}
\end{align}
the Deligne-Mumford compactification of 
the moduli space $M_{0,4}$ of $4$-pointed projective lines.
It is one-dimensional, because the first three punctures can
be transformed to $\mathset{0,1,\infty}$, whereas the fourth
puncture runs through $\lambda\in\mathbb{P}^1\setminus\mathset{0,1,\infty}$.
The boundary is given by
letting $\lambda$ run into $\mathset{0,1,\infty}$. In order
to also have $4$ punctures in this case, one takes  the singular curves
as depicted in Figure \ref{stable4dendro}. These are
stable $4$-pointed genus zero curves. 

\begin{figure}[h]
$$
\begin{array}{c}
\xymatrix@=3pt{
&&\infty&&\\
 A\colon&&*\txt{$\bullet$}\ar@{-}[u]\ar@{-}[dl]\ar@{-}[ddrr]&&\\
&*\txt{$\bullet$}\ar@{-}[dl]\ar@{-}[dr]&&&\\
0&&1&&\lambda}
\end{array}
\hfill
\begin{array}{c}
\xymatrix@=3pt{
&&\infty&&\\
 B\colon&&*\txt{$\bullet$}\ar@{-}[u]\ar@{-}[dl]\ar@{-}[ddrr]&&\\
&*\txt{$\bullet$}\ar@{-}[dl]\ar@{-}[dr]&&&\\
0&&\lambda&&1}
\end{array}
\hfill
\begin{array}{c}
\xymatrix@=3pt{
&&\infty&&\\
 C\colon&&*\txt{$\bullet$}\ar@{-}[u]\ar@{-}[dl]\ar@{-}[ddrr]&&\\
&*\txt{$\bullet$}\ar@{-}[dl]\ar@{-}[dr]&&&\\
1&&\lambda&&0}
\end{array}
\hfill
\begin{array}{c}
\xymatrix@=10pt{
&\infty&\\
D\colon&*\txt{$\bullet$}\ar@{-}[u]\ar@{-}[dl]\ar@{-}[d]\ar@{-}[dr]&\\
0&1&\lambda
}
\end{array}
$$
\caption{Trees representing the different cells of ${M}_{0,4}$.}
\label{d3}
\end{figure}
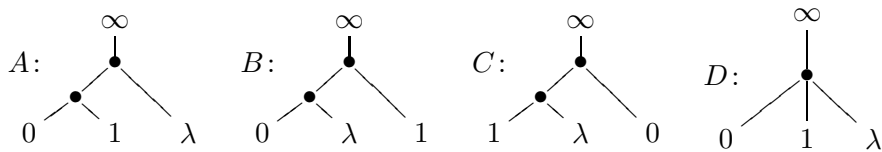

\begin{figure}[h]
$$
\xymatrix@R=3pt@C=5pt{
&&&&\\
&&&&\\
&&&&\\
&&&*\txt{$\bullet$}
\ar@{-}[uuu]_{A}\ar@{-}[ddll]_{B}\ar@{-}[ddrr]^{C}&\\
&&&\scriptstyle D&\\
&&&&&&\\
}
$$ 
\caption{The cell structure of $M_{0,4}^{\trop}$.}
\label{M04trop}
\end{figure}
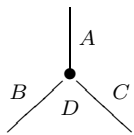

From (\ref{P1=M04}), 
we can also write the $4$-point amplitude (\ref{4ptstringamp}) as
$$
A_p^0(k_1,k_2,k_3,k_4)=
\int_{\bar{M}_{0,4}}\absolute{x}^{k_1\cdot k_2}
\absolute{1-x}^{k_1\cdot k_3}\;dx,
$$
and consider now the contributions from different parts of the moduli space. 
By looking at the possible
trees $\mathscr{T}(0,1,\infty,\lambda)$
depicted in Figure \ref{d3},
we see that the moduli space $M_{0,4}$ decomposes into
4 cells $A,B,C,D$. The first three cells which allow the edge length
in the tropical curve
to vary, are homeomorphic to the open unit intervall, whereas
cell $D$ is $0$-dimensional. The corresponding 
cell structure of the moduli space of tropical curves is illustrated in
Figure \ref{M04trop}. 

Observe that $D$ looks like a zero set. Indeed,
similarly as in  Section \ref{smoothy},  
we can find a 
measure on $\mathbb{P}^1\cong\bar{M}_{0,4}$ 
which induces the uniform distribution
$d\lambda$
on $\bar{M}_{0,4}^{\trop}$ via the tropicalisation map
$$
\trop\colon\mathbb{P}^1\to\bar{M}_{0,4}^{\trop}.
$$
With $g(x)=\absolute{x}^{k_1\cdot k_2}\absolute{1-x}^{k_1k_3}$,
it follows that 
$$
\int_{D}g c_1(\bar{O}(1))=\int_D c_1(\bar{O}(1))
=\int_{\trop(D)} d\lambda=0,
$$
where the first equality\footnote{In \cite{Gubler2007},
the notation $\int f c_1(\bar{O}(1))$ is preferred
to our $\int g c_1(\bar{O}(1))$, where
$f=-\log g$.}  follows from  $g|_D=1$.

\medskip
This approach  generalises to the case of $n$ points. Namely, 
assume we are given vectors
$k_1,\dots,k_n\in\mathbb{C}^d$ with $\sum
_{i=1}^nk_i=0$
and $k_i^2=2$. Then we obtain:

\begin{thm} \label{zeroconj}
The $p$-adic $n$-point tachyon string  amplitude at the tree level
\begin{align*}
A^0_p&(k_1,\dots,k_n)
\\&=\int\limits_{\bar{M}_{0,n}}dx_2\dots dx_{n-2}
\prod\limits_{i=2}^{n-2}\absolute{x_i}^{k_1\cdot k_i}
\absolute{1-x_i}^{k_{n-1}\cdot k_i}
\prod\limits_{2\le i<j\le n-2}
\absolute{x_i-x_j}^{k_i\cdot k_j}
\end{align*}
is contributed
in the tropical limit only by those kinds of $n$-point configurations
$\mathbb{P}^1\setminus\mathset{0,1,x_2,\dots,x_{n-2},\infty}$
for which $\mathscr{T}(0,1,x_2,\dots,x_{n-2},\infty)$
is a binary tree.
\end{thm}

\section{Counting plane curves}

Here, we sketch a construction of $p$-adic line bundles
on the moduli space $\bar{M}_{g,n}$ of stable $n$-pointed genus $g$
curves which can be used in order to count curves in the plane 
passing through given points satisfying some tangency conditions.
Our approach is similar to the one in the previous section,
except that
 now we consider the case in which the physics is ``removed''
from the problem. 

\subsection{$p$-adic $\psi$-classes?}

The idea of $\psi$-classes is to allow the counting of curves with prescribed tangency conditions
when passing through some prescribed subspaces of some target space.

Let $L_i$ be the line bundle on $\bar{M}_{g,n}$ which yields in every curve $C$ 
represented by $x=[C,p_1,\dots,p_n]\in\bar{M}_{g,n}$ the cotangent in $p_i$.
This is called the {\em $i$-th cotangent bundle} on $\bar{M}_{g,n}$.
In complex algebraic geometry,
the $\psi$-class $\psi_i$ is then defined as 
a certain cohomology class called the {\em first Chern class}
of $L_i$:
$$
\psi_i:=c_1(L_i)\in H^2(\bar{M}_{g,n},\mathbb{Z})
$$
which is nothing but the algebraic curvature encountered
already in Section \ref{smoothy}.
One obtains an intersection product
\begin{align}
\langle\tau_{k_1}\cdots\tau_{k_n}\rangle:=\int_{\bar{M}_{g,n}}\psi^{k_1}\wedge\dots\wedge
\psi^{k_n}\label{GWdesc}
\end{align}
which takes non-zero values if and only if $\sum k_i=3g-3+n$.

The notation $\langle\tau_{k_1}\dots\tau_{k_n}\rangle$
introduced by Witten \cite{Witten}
suggests a ``physical'' interpretation of the $\tau_{k_i}$
as operators on some Hilbert space whose correlator is the integral.
In any case, the value of $\langle\tau_{k_1}\cdots\tau_{k_n}\rangle$
is symmetric in $k_1,\dots,k_n\in\mathbb{N}$.
The expression $\psi_1^{k_1}\wedge\dots\wedge\psi_n^{k_n}$ can also be seen as a positive
measure $\mu$ on the space $\bar{M}_{g,n}$ with total mass $\mu(\bar{M}_{g,n})$
given by (\ref{GWdesc}).

\smallskip
Unfortunately, there is no sensible $p$-adic notion of Chern class of line bundles.
The consequence is that there are no $p$-adic $\psi$-classes at hand. However,
the $p$-adic analogon of the measure $\mu$ can be constructed
as in Section \ref{smoothy}.
Namely, take an $O_K$-model of $\bar{M}_{g,n}$ whose special fibre is a blow up
of $\mathcal{M}:=\bar{\mathcal{M}}_{g,n}\otimes \kappa$ in the boundary in such a way that the
vertices of $\bar{M}_{g,n}^{\trop}$ correspond to the irreducible components of $\mathcal{M}$.
Take an
$O_K$-model $\mathscr{L}_i$ of $L_i$.
Then  (\ref{CL-measure})
defines a measure 
$$
c_1(\bar{L}_1)^{k_1}\wedge\dots\wedge c_1(\bar{L}_n)^{k_n}
$$
on $\bar{M}_{g,n}$ which
 is supported on the points above the generic points of the components
of $\mathcal{M}$.
A smoothing process yields as in Section \ref{smoothy} a measure $\mu_p$ 
for which $\trop_*(\mu_p)$ 
is a piece-wise Haar measure on $\bar{M}_{g,n}^{\trop}$ \cite{Brad}.
It is uniform on the closure of each maximal cell of $\bar{M}_{g,n}^{\trop}$.
Now,
each cell parametrises tropical curves of fixed combinatorial type,
and the maximal cells correspond to trivalent semi-graphs.
So, we state our result:
\begin{thm}
The $p$-adic ``correlator''
$$
\langle\tau_{k_1}\cdots\tau_{k_n}\rangle:=\int_{\bar{M}_{g,n}}\mu_p
$$
is contributed only by the locus of trivalent Mumford curves in 
$\bar{M}_{g,n}$, and is a weighted graph sum.
\end{thm}

The proof follows by observing that
$
\trop_*(\mu_p)
$
is a measure for which all cells of dimension lower than $3g-3+n$
are zero sets. 

\subsection{Including ``gravity''}

We now consider the problem of counting curves of degree $d$ and genus $g$
passing through $n=3d+g-1$ points in the plane. A solution to this problem
was predicted through Witten's conjecture \cite{Witten}, proved by Kontsevich
\cite{KM1994}. The idea is to count maps $C\to\mathbb{P}^2$ of $n$-pointed curves
into the plane, called ``instantons''. The existence of a target space $X$ (here: $\mathbb{P}^2$)
introduces ``gravity'' to the system. By using so-called {\em stable maps},
one obtains a compactification of the moduli space of instantons.
The theory then allows the construction of ``gravitational'' $\psi$-classes
and correlators.

Recently, it was shown that counting maps of tropical curves to the tropical
plane $\mathbb{TP}^2$ yields the same numbers as for usual curves \cite{MikhalkinJAMS,GathmannMarkwigWDVV}.
Those numbers are also called {\em Gromov-Witten invariants}.
From a $p$-adic point of view, the correspondence between the classical and
tropical Gromov-Witten numbers does not come as a surprise. Namely, we
have a commuting diagram
$$
\xymatrix{
C\ar[d]_{\trop}\ar[r]&\mathbb{P}^2\ar[d]^{\Val_{\mathfrak{A}}}\\
\Gamma\ar[r]&\mathbb{TP}^2
}
$$
with a lot of choices of maps $\Val_{\mathfrak{A}}\colon\mathbb{P}^2\to\mathbb{TP}^2$.
Namely, for any configuration $\mathfrak{A}$ of three lines in $\mathbb{P}^2$
in general position, there is a transformation 
$\alpha\in\PGL_3(K)$ which takes $\mathfrak{A}$ to the three standard lines,
i.e.\ the 2 coordinate lines in $K^2$ and the line at infinity.
Then define
$$
\Val_{\mathfrak{A}}:=\Val\circ\alpha.
$$
Classically, the number of curves passing through 
a set $\mathscr{P}$ of $n=3d+g-1$ points
in $\mathbb{P}^2$ does not depend on the position of the points,
as long as they are in general position.
It follows that if for some $\mathfrak{A}$, the set
$\Val_{\mathfrak{A}}(\mathscr{P})$ consists of $n$ points
in $\mathbb{TP}^2$ tropically in general position, then the number of tropical
curves $\Gamma$ with $b_1(\Gamma)=g$ and degree $d$ passing through
$\Val_{\mathfrak{A}}(\mathscr{P})$ does not depend on their positions
in $\mathbb{TP}^2$.
As a side effect of this observation, we obtain the result:

\begin{thm}
If there exists a line configuration $\mathfrak{A}$ such that
$\Val_{\mathfrak{A}}(\mathscr{P})$ consists of $n$ points tropically
in general position, then 
$$
N_{d,g}^{\Mumf}(\mathscr{P})=N_{d,g}.
$$
i.e.\ the plane curves of degree $d$ and genus $g$ passing through $\mathscr{P}$
are all Mumford curves.
\end{thm}


\subsection{The CMZ-conjecture in adelic string theory}

A crucial observation in Section \ref{ptreeamp}
 was that integrating over all $n$-point configurations 
on the projective line
means 
in fact integration over the moduli space $M_{0,n}$ of $n$-pointed
genus $0$ curves. Hence, a straightforward generalisation to the multiloop 
case means to integrate over $M_{g,n}$, the moduli space of
$n$-pointed genus $g$ curves, resp.\ its Deligne-Mumford compactification 
$\bar{M}_{g,n}$.
Indeed, Chekhov et al.\ \cite{CMZ1989} describe a $p$-adic multiloop
amplitude. However, in their calculations
they vary only the $n$ points on the $p$-adic Riemann surface $X$
while  keeping the surface itself fixed.
But their conjecture \cite[Conj.\ 4.3.3]{CMZ1989} is a statement about the amplitude when
both, the points and the holomorphic structure on $X$, vary.
Tropically, this amounts to varying the possible combinatorial types
of $\Gamma=\trop X$ as well as the possible lengths of the bounded edges of $\Gamma$.
Hence, we can formulate:

\begin{conj}
The $p$-adic string amplitude
$$
A^g_p(k_1,\dots,k_n)=
\int_{\bar{M}_{g,n}^{\trop}} \trop_*\mu_p
$$
is  contributed 
in the tropical limit
by
precisely the binary Mumford curves
via weighted summation of the graphs underlying their tropicalisations. 
\end{conj}

\section*{Acknowledgements}

The author 
would like to thank Branko Dragovich for the invitation to give a lecture
in the 5th Summer School and Conference on Modern Mathematical Physics
in Belgrade (July 2008),
and for pointing out errors in a first draft
of this article. The work reported was partially supported by
the JSPS Postdoctoral Fellowship Programme (short term) in  research project
PE 05573, and also from  DFG-project BR 3513/1-1. He thanks Fumiharu Kato
for  fruitful discussions  and kind hospitality 
at Kyoto University, 
as well as Fionn Murtagh for directing 
the author's attention 
to the review  article \cite{BF1993}  referring to  the CMZ-conjecture.

\end{document}